\documentclass[prl,aps,superscriptaddress,onecolumn]{revtex4-1}
\usepackage{graphicx} 
\graphicspath{{graph/}}
\usepackage[usenames]{color}
\usepackage{amsmath,amssymb}
\usepackage{gensymb}
\usepackage{natbib}
\usepackage{xcolor}
\def\strutdepth{\dp\strutbox}
\def\nw#1{\strut\vadjust{\kern-\strutdepth\vtop to0pt{\vss\hbox to\hsize
{\hskip\hsize\hskip5pt$\leftarrow$\hss\strut}}}{\em #1}}

\usepackage{rotating}
\usepackage{multirow}

\begin{document}

\title{The paradox of contact angle selection on stretched soft solids}
\author{Jacco H. Snoeijer}
\affiliation{Physics of Fluids Group, Faculty of Science and Technology, University of Twente, P.O. Box 217, 7500 AE Enschede, The Netherlands}
\affiliation{Department of Applied Physics, Eindhoven University of Technology, P.O. Box 513, 5600MB, Eindhoven, The Netherlands}
\author{Etienne Rolley}
\affiliation{Laboratoire de Physique Statistique, UMR 8550 ENS-CNRS, Univ. Paris-Diderot, 24 rue Lhomond, 75005, Paris.}
\author{Bruno Andreotti}
\affiliation{Laboratoire de Physique Statistique, UMR 8550 ENS-CNRS, Univ. Paris-Diderot, 24 rue Lhomond, 75005, Paris.}

\begin{abstract}
The interfacial mechanics of soft elastic networks play a central role in biological and technological contexts. Yet, effects of solid  capillarity have remained controversial, primarily due to the strain-dependent surface energy. Here we derive the equations that govern the selection of contact angles of liquid drops on elastic surfaces from variational principles. It is found that the substrate's elasticity imposes a nontrivial condition that relates pinning, hysteresis and contact line mobility to the so-called Shuttleworth effect. We experimentally validate our theory for droplets on a silicone gel, revealing an enhanced contact line mobility when stretching the substrate. 
\end{abstract}
 
\date{\today}

\maketitle

The functionality of extremely soft materials emerges from a combination of bulk elasticity and surface effects \cite{RB2010,Mora:2010aa,AS16,SJHD2017,BRR2018}. However, the interfacial mechanics of soft solids, typically reticulated polymer networks, is notoriously difficult to probe experimentally \cite{NadermannPNAS2013,MondalPNAS2015}. A very promising route to quantitatively measure solid surface tension is via the contact angles of liquid drops \cite{Style13,XuNatComm2017,Schulman2018}. The wetting on soft solids is intermediate between the case of rigid solids for which the contact angle is selected by Young's law, and the case of a liquid-liquid interface for which contact angles are selected by Neumann's law \cite{Style2012a,MDSA12b,Lub14,DL15,MDNeumann}. However, the interpretation of contact angles on stretched solids has recently raised a controversy, with similar experiments leading opposite conclusions on the coupling between elasticity and surface tension~\cite{XuNatComm2017,Schulman2018}. In this Letter, we resolve this paradox by deriving the equilibrium conditions at the contact line from first principles. We reveal a previously ignored condition that must be satisfied to avoid pinning and contact angle hysteresis. Predictions are validated by dynamical experiments \cite{Carre1996a,LAL96,Kajiya2013a,Dupas2014a,KarpNcom15}, elucidating how stretching the substrate affects hysteresis and contact line mobility.

The challenge arises due to the fundamental difference between the capillarity of solids and liquids. For solid interfaces, the excess energy $\gamma$ per unit area generically depends on the surface strain $\epsilon$. Contrarily to liquid interfaces, the surface tension $\Upsilon$ in the interface therefore differs from $\gamma$: surface energy and surface tension are related by the Shuttleworth equation~\cite{Shut50,AS16,SJHD2017}, 
\begin{equation}
\Upsilon(\epsilon)=\frac{d }{d \epsilon}\left[ (1+\epsilon)\gamma(\epsilon) \right]=\gamma+(1+\epsilon) \gamma'.
\label{Shuttleworth}
\end{equation}
The first evidence of a strong ``Shuttleworth effect" for reticulated polymers was obtained using an elastic (polyvinylsiloxane) Wilhelmy plate, which allows for the measurement of the difference $\gamma'_{SL}-\gamma'_{SV}$ between solid/liquid and solid/vapour interfaces~\cite{Marchand2012c,AS16}. Recent studies addressed the Shuttleworth effect through contact angles of liquid drops (cf. Fig.~\ref{Fig1}), in particular their variation when stretching the substrate \cite{XuNatComm2017,Schulman2018}. Intriguingly, the observations led to contradictory interpretations. Xu \emph{et al.}~\cite{XuNatComm2017} observed that stretching a silicone gel leads to a significant increase of the \emph{solid} angle $\theta_S$, which was attributed to a strong Shuttleworth effect $\gamma'$. Schulman \emph{et al.}~\cite{Schulman2018}, by contrast, conclude that there is no Shuttleworth effect for a broad range of different elastomers. This is based on the striking absence of any dependence of the \emph{liquid} angle $\theta_L$: While the contact angle on (stiff) glassy polymers varies with the external strain, no variation of $\theta_L$ was found for (soft) elastomers up to 100\% strain. These observations point to a pressing lacuna in the understanding of solid capillarity. 
\begin{figure}[t!]
\includegraphics{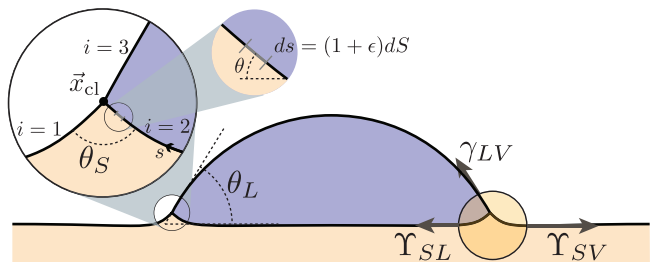}
\vspace{- 4 mm}
\caption{Contact angles on a soft solid: The liquid angle $\theta_L$ of the drop (main panel) and the solid angle $\theta_S$ of the wetting ridge (zoom). In the presented theory, the three interfaces ($i=1,2,3$) are parametrised using curvilinear coordinates $s$, related to the reference coordinate $S$ by the strain $\epsilon_i(S)$, while the local slope of the interface is $\theta_i$(S). The interfaces meet at the contact line position $\vec x_{\rm cl}$. The effect of surface tensions $\Upsilon_{SV}$ and $\Upsilon_{SL}$ is illustrated by a force balance on the circular zone near the contact line on the right [Eq.~(\ref{BalanceTripartite},\ref{deGennes})].}
\label{schematics}
\vspace{- 5 mm}
\label{Fig1}
\end{figure}

{\it Exact wetting conditions from variational analysis~--~} Here we set out to derive the complete set of equilibrium conditions for soft wetting, by simultaneously minimising elastic and capillary energies. The geometry is sketched in Fig.~\ref{Fig1}, showing the three interfaces near the contact line. The index  $i=1$ refers to the solid/vapor interface, $i=2$ refers to the solid/liquid interface and $i=3$ refers to the liquid/vapor interface. From the outset we consider that the size of elastic deformation $\sim \gamma/G$ is large compared to molecular scales ($G$ being the static shear modulus), so that a wetting ridge develops and the interfaces are sharp. The free energy per unit width reads
\begin{equation}
\mathcal{F} =  \mathcal{F}_e + \sum_i \int_{-\infty}^{r_i} \gamma_i(\epsilon_i)  \,  ds,
\end{equation}
where $\mathcal F_e$ is the elastic energy, and $s$ is the curvilinear coordinate along each of the interfaces. The integrals runs from a position far from the contact line (``-$\infty$") to the contact line ($s=r_i$). The strain field $\epsilon_i$ is defined for the elastic interfaces ($i=1,2$) and is actually a function of the reference coordinate $S$. This coordinate refers to a material point at the interface in its undeformed state prior to the deposition of the liquid drop. The geometric connection between the deformed state and reference state (Fig.~\ref{schematics}) reads $\gamma_i(\epsilon_i)  \,  ds=  (1+\epsilon_i) \gamma_i(\epsilon_i)  \,  dS$, so that the variation $\delta \epsilon_i(S)$ naturally gives rise to $\Upsilon_i$ as defined in~(\ref{Shuttleworth}). The interfaces are fully specified when complementing the strain with the local angle $\theta_i(S)$, which must also be varied to obtain the equilibrium. Minimization of $\mathcal F$ then implies $\delta \mathcal F_{e}/\delta  \epsilon_i + \Upsilon_i = 0$ and $\delta \mathcal F_{e}/\delta  \theta_i=0$.

It is convenient to express the derivatives of the elastic energy in terms of the elastic traction $\vec \sigma$, defined as the elastic force per unit (deformed) area. It can be inferred from the variation of interface displacements $\delta \vec u$, as derived in the Supplementary Information~\cite{sup}. From the kinematic connection between $\delta \epsilon$, $\delta \theta$ and $\delta \vec u$, one finds (omitting the indices $i=1,2$)
\begin{eqnarray}\label{eq:traction}
\vec \sigma(s) = \frac{\delta \mathcal{F}_{e}}{\delta \vec u(s)} = 
- \frac{d}{ds}\left( \frac{\delta \mathcal F_e}{\delta \epsilon(S)} \vec t + \frac{1}{1+\epsilon} \frac{\delta \mathcal F_e}{\delta \theta(S)}   \vec n \right).
\end{eqnarray} 
Here $\vec t$ and $\vec n$ respectively are tangential and normal unit vectors along the interfaces. Combined with the mentioned conditions for $\delta \mathcal F=0$, the traction (\ref{eq:traction}) becomes
\begin{equation}\label{eq:marangoni}
\vec \sigma(s)=\frac{d \left( \Upsilon \vec t\right)}{ds}=\frac{d \Upsilon}{ds} \vec t+ \Upsilon \frac{d\theta}{ds}\vec n.
\end{equation}
The elastic traction $\vec \sigma$ balances the tangential Marangoni stress due to gradients of surface tension and normal Laplace pressure due to curvature, as in liquid capillarity.

\begin{figure}[t!]
\includegraphics{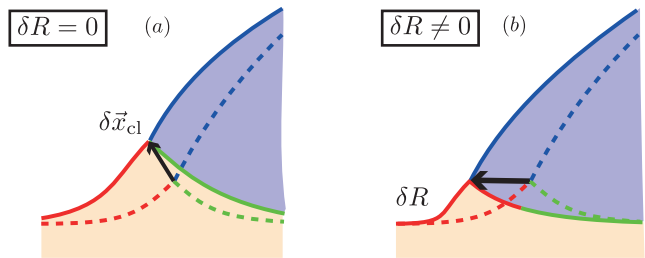}
\vspace{- 4 mm}
\caption{Equilibrium requires that the energy is stationary with respect to contact line displacements. (a) Displacement at fixed material point ($\delta R=0$), where the contact line remains pinned to the solid. (b) The displacement with variable material point ($\delta R \neq 0$) gives rise to a \emph{no-pinning condition}. Color indicates whether material points in the reference state belonged to the wet part (green) or dry part (red) of the solid.}
\label{DisplacementCL}
\vspace{- 5 mm}
\label{Fig2}
\end{figure}

The key purpose of the analysis, however, is to derive the \emph{boundary conditions} at the contact line. For this we need to specify how the three interfaces $i=1,2,3$ are connected at their respective end points at $S=R_i$. Obviously, the interfaces  meet at a common position $\vec x_{\rm cl}$ (Fig.~\ref{schematics}), which must be imposed as a constraint. The constraint does not affect (\ref{eq:marangoni}), but varying the contact line position $\delta \vec x_{\rm cl}$ at constant $R_i$, gives a boundary condition evaluated at the contact line (Suppl. Inf.~\cite{sup}):
\begin{equation}\label{eq:neumann}
\sum \Upsilon_i \vec t_i=\vec 0.
\end{equation}
This is the Neumann law that determines $\theta_S$, commonly used but here derived from variational principles. Importantly, Eq.~(\ref{eq:neumann}) is obtained by ``pinning" the contact line to a fixed material point of the elastic interface, i.e. $S=R_1$ and $S=R_2$ were kept constant, as sketched in Fig.~\ref{Fig2}a. Without contact angle hysteresis, however, there can be no such pinning: As is illustrated in Fig.~\ref{Fig2}b, the liquid can freely move and change the solid molecules that are present at $\vec x_{\rm cl}$. This must be accounted for by allowing the variation $\delta R_1 = -\delta R_2$, exchanging material points from the dry ($i=2$) to the wetted interface ($i=1$). The variation $\delta R$ gives a new boundary condition at the contact line, which will be referred to as the \emph{no-pinning condition}:
\begin{equation}
\Delta\left[(1+\epsilon)^2\gamma'(\epsilon)- \frac{\partial \mathcal F_e}{\partial R} \right]=0. \label{eq:SecondBC}
\end{equation}
Here $\Delta[\cdots ]$ denotes the difference between both sides of the contact line, and controls the presence of discontinuities of stress and/or strain (Suppl. Inf.~\cite{sup}). Although the appearance of two boundary conditions (\ref{eq:neumann}) and (\ref{eq:SecondBC}) is logically associated with the position $\vec x_{\rm cl}$ and the material point of the solid $R_i$, equation (\ref{eq:SecondBC})  had never been considered so far. Note that for rigid solids, the variation $\delta x_{\rm cl}$ is automatically accompanied by a change of the material point $R_i$ so that it gives a no-pinning condition.

In order to compare to macroscopic experiments, we need to translate these results to regions far away from the wetting ridge (Fig.~\ref{Fig1}). To this end we first integrate eq.~(\ref{eq:marangoni}) across the contact line, which using eq.~(\ref{eq:neumann}) gives 
\begin{eqnarray}
\Upsilon_{SL}-\Upsilon_{SV}+\gamma_{LV} \cos \theta_L&=&\vec e_x \cdot \int_{-\infty}^\infty  \vec \sigma ds  \label{BalanceTripartite}\\
\gamma_{LV} \sin \theta_L&=&\vec e_y \cdot \int_{-\infty}^\infty  \vec \sigma ds \label{deGennes}.
\end{eqnarray}
This expresses the resultant force of the three surface tensions on the volume indicated in Fig.~\ref{Fig1}, which must be balanced by the elastic stress integrated over the contact line region. Besides the normal force, we recover the appearance of a tangential stress across the contact line \cite{DMAS2011,Marchand2012c,AS16}. Recombining these equations with the no-pinning condition (\ref{eq:SecondBC}) this can be expressed as
\begin{eqnarray}
(1+\epsilon_\infty)\left( \gamma_{SL}  -  \gamma_{SV} +\gamma_{LV} \cos \theta_L \right) = - \Delta \left[ (1+\epsilon)^2 \gamma'\right] \nonumber \\
+ \int_{-\infty}^\infty  \vec \sigma \cdot \left[ (1+\epsilon) \vec t - (1+\epsilon_\infty) \vec e_x \right] ds. \quad
\label{YoungBalance}
\end{eqnarray}
%
This is the generalisation of Young's law for the liquid angle $\theta_L$ far away from the contact line, also applicable when the substrate is uniaxially strained to a value $\epsilon_\infty$.

{\it The paradox~--~} To resolve the experimental controversy regarding the Shuttleworth effect, we evaluate these exact results in the framework of linear elasticity for which $|\epsilon| \ll 1$. The elastic energy is then obtained by the surface integral $\mathcal F_e= \int ds \, \frac{1}{2}\vec u \cdot  \vec \sigma$. Hence, the boundary condition (\ref{eq:SecondBC}) then simplifies to $\Delta\left[\gamma'(0)- \frac{1}{2} \vec u \cdot  \vec \sigma \right]=0$. A stress discontinuity is not admitted in linear elasticity, as it would lead to a logarithmic singularity of slope. Hence, we deduce that the newly found no-pinning condition  (\ref{eq:SecondBC}) enforces continuity of $\gamma'(0)$, so that $\gamma_{SV}'=\gamma'_{SL}$. Far from the contact line this implies that (\ref{YoungBalance}) reduces to the standard Young equation $ \gamma_{SL}  -  \gamma_{SV} +\gamma_{LV} \cos \theta_L=0$. All the above remains valid for large $\epsilon_\infty$, as long as the perturbation induced by the deposited drop remains in the regime of linear response.

This explains why $\theta_L$ \emph{must} remain constant on stretched elastomers \cite{Schulman2018}, even when the Shuttleworth effect leads to a change of $\theta_S$~\cite{XuNatComm2017}. Namely, the no-pinning condition $\gamma_{SV}'=\gamma'_{SL}$ imposes that the difference $\gamma_{SV}-\gamma_{SL}$ stays constant. Hence, as long as the drop's distortion of the solid is within linear response, the constancy of $\theta_L$ is a direct consequence of the lack of hysteresis. This is a very surprising result since it is not clear a priori why $\gamma_{SV}'$ would equal $\gamma'_{SL}$ for generic polymer networks. Failure of this equality must either result into a ``process zone" of large strains near the contact line to satisfy the no-pinning condition (\ref{eq:SecondBC}), or otherwise pinning must be observed. Indeed, elastomeric surfaces can exhibit a strong contact angle hysteresis, which we suggest finds its origin in the Shuttleworth effect. For example, the Wilhelmy plate experiment of~\cite{Marchand2012c} for which strong hysteresis was present, exhibited $\gamma_{SL}'\ne \gamma_{SV}'$ as inferred from the tangential stress balance (\ref{BalanceTripartite}). 


%
%

{\it Contact angle selection from dynamic spreading~--~}We now experimentally validate this interpretation framework: Linear response and absence of hysteresis necessarily implies the equality $\gamma_{SV}'=\gamma_{SL}'$ -- and hence $\theta_L$ independent of strain. A proper determination of hysteresis calls for dynamic spreading experiments, where the contact angle is measured versus contact line velocity $v$~\cite{PLDRA2016}. The hysteresis is then inferred from the limit of vanishing velocity, comparing the advancing motion ($v \rightarrow 0^+$) and the receding motion ($v \rightarrow 0^-$). Here we use the same set up as in \cite{KarpNcom15} to measure the macroscopic contact angle $\theta_L$, adapted to impose a uniaxial strain $\epsilon_\infty$ to a PDMS gel (Dow Corning CY52-276). A liquid droplet is inflated or deflated with a syringe to impose an advancing  or a receding motion. Milli-Q water and fluorosilicone oil (poly(3,3,3-trifluoropropylmethylsiloxane), Gelest FMS 121) were used.
\begin{figure}[t!]
\includegraphics{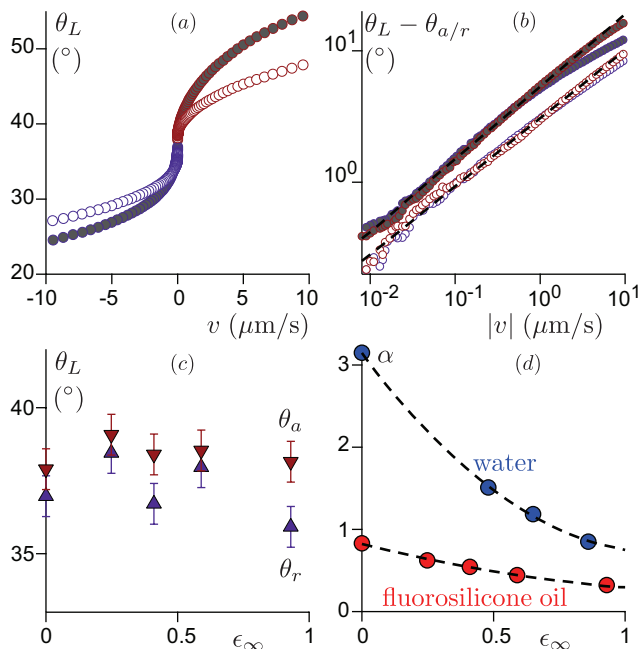}
\vspace{- 4 mm}
\caption{Pinning and enhanced contact line mobility. (a) Liquid-vapor macroscopic contact angle $\theta_L$ with respect to the undeformed solid surface as a function of the contact line velocity $v$ (liquid: fluorosilicone oil).  Closed and open symbols: un-strained and strained ($\epsilon_\infty=0.93$) samples respectively. (b)  Contact angle rotations $\theta_L-\theta_{a/r}$ vary like $v^n$, with $n$ obtained from loss modulus measurement as expected from (\ref{eq:dynamics}) ($n=0.55$ for $\epsilon_\infty=0$ and $n=0.50$ for $\epsilon_\infty=0.93).$ (c) Liquid contact angles $\theta_{a/r}$ in the limit $v \rightarrow 0^+ / 0^-$, as a function of the applied strain $\epsilon_\infty$.  (d) Contact line friction factor $\alpha$ defined by (\ref{eq:dynamics}) as a function of strain $\epsilon_\infty$.} 
\vspace{- 5 mm}
\label{ThetaDynamics}
\end{figure}

The resulting liquid contact angle $\theta_L$  is shown in Fig.~\ref{ThetaDynamics}a as a function of the contact line velocity $v$, for unstrained (closed circles) and strained samples (open circles). The curves apparently exhibit no discontinuity at $v=0$ pointing to a nearly perfect absence of hysteresis. Both the advancing and receding motions exhibit a power-law dependence $\theta_L - \theta_{a/r} \sim |v|^n$ over more than two decades in velocity (Fig.~\ref{ThetaDynamics}b). This allows us to determine accurately the advancing $\theta_a$ and receding  $\theta_r$ angles, revealing a small ($\sim 1^\circ$) constant hysteresis (Fig.~\ref{ThetaDynamics}c). Furthermore, the equilibrium liquid angle $\theta_a \simeq \theta_r$ is independent of the imposed strain $\epsilon_\infty$, consistently with previous independent experiments \cite{XuNatComm2017,Schulman2018}. The condition $\gamma'_{SL}=\gamma'_{SV}$, predicted by our theory in the absence of hysteresis, is therefore fulfilled in our experiment. 

Another remarkable result of Fig.~\ref{ThetaDynamics} is that the spreading velocities are strongly enhanced upon stretching the solid. Namely, comparing the data for strained (open) and unstrained (closed) samples for a given $\theta_L$,  the velocity $|v|$ is larger by a factor $3$. Hence, we observe that stretching leads to an enhanced wetting mobility. 

The final step is to experimentally verify our assumption of linear response and to explain the enhanced wetting mobility on a stretched gel. The gel's linear rheology under uniaxial strain turns out to only weakly depend on the imposed $\epsilon_\infty$~(Suppl. Inf.~\cite{sup}). The loss modulus $G''$ depends as a power law of the angular frequency $\omega$: $G'' \propto G (\omega \tau)^n$, where the cross-over time $\tau$ and the exponent $n$ marginally depend on $\epsilon_\infty$ (Suppl. Inf. \cite{sup}). This enables us to use the dynamical theory from~\cite{KarpNcom15}, relating $\theta_L$ to the contact line velocity $v$ 
\begin{equation}\label{eq:dynamics}
\theta_L-\theta_{a/r} =  \alpha\, \left(\frac{G\,|v|\tau}{\gamma_{LV} \sin \theta_L}\right)^{n},
\end{equation}
where the dimensionless friction factor $\alpha$ depends on the geometry of the ridge. Indeed, the exponents $n$ measured for $\theta_L$ and in the linear rheological measurement are found to be consistent, as predicted by~(\ref{eq:dynamics}). The agreement of these exponents provides a direct proof that the droplet dynamics probes the substrate within linear response, even when the pre-strain $\epsilon_\infty$ is not small. 

The experimentally measured friction factor $\alpha$ defined by (\ref{eq:dynamics}) is reported as a function of $\epsilon_\infty$ in Fig.~\ref{ThetaDynamics}d. The reduction of friction with strain can be attributed to the Shuttleworth effect, via a gradual increase of $\theta_S$: A shallower wetting ridge leads to a smoother motion of material points and hence to less dissipation. This effect can be estimated from viscoelastic theory~\cite{KarpNcom15} (based on constant surface tensions), suggesting the scaling
\begin{equation}
\alpha \sim  \cos^{1+n} (\theta_S/2)
\end{equation}
Hence, the reduced friction $\alpha$ in our spreading experiments (Fig.~\ref{ThetaDynamics}d) points to a Shuttleworth-induced gradual increase of $\theta_S$ with $\epsilon_\infty$ -- consistently with direct measurements of $\theta_S$ in~ \cite{XuNatComm2017}. Obviously, an important step for future work is to achieve fully self-consistent computations of the ridge mechanics that includes the strains induced by the Shuttleworth effect. This would e.g. lead to a fully quantitative prediction of $\alpha(\theta_S)$.

{\it Outlook~--~}Our study offers a general framework that establishes the laws of wetting on deformable soft solids. We have shown that equilibrium, in particular the lack of hysteresis, requires the equality of Shuttleworth coefficients $\gamma'_{SL}=\gamma'_{SV}$ for both the wet and dry solid. This sheds an unexpected light on coupling of physical chemistry, encoded in $\gamma(\epsilon)$, to the mechanics of wetting, and calls for a better understanding of the molecular origin of the Shuttleworth effect in cross-linked polymers. This is of prime importance for the design and rheological characterisation of extremely soft materials, for which the interfacial effects dominate over bulk elasticity. From the perspective of wetting applications, we demonstrate that the Shuttleworth effect offers a new route to control the mobility of the contact line as illustrated here by stretching-enhanced spreading velocities. 

 This work was financially supported by the ANR grant Smart and ERC (the European Research Council) Consolidator Grant No. 616918.

\newpage
{\bf \Large Online supplement\\}
In this supplement we present some technical details regarding the derivation of the theory -- in particular the treatment of the constraints. This is complemented with details of the rheological characterisation of the strained samples. \\

\section{Kinematics}
\subsection{From curvilinear to Cartesian}
We specify how the curvilinear parametrisation defined in the paper is related to Cartesian coordinates. This is necessary to specify the positional constraint at the contact line and the connection to elastic stress (Eq. 3 in the main text). As is recalled in Fig.~\ref{Fig1}, we introduce three interfaces labelled by $i=1,2,3$, each of which is characterized by a reference coordinate system $S_i \in [-\infty,R_i]$. We introduce a strain field $\epsilon_i$ at the interface, such that 
\begin{equation}
ds_i = dS_i (1+\epsilon_i),
\end{equation}
where the $s_i$ are the curvilinear coordinates in the current state (after stretching). The shape is fully specified when complemented by the local interface angle $\theta_i(S_i)$. In the following we omit the subscript $i$ for simplicity. The connection to Cartesian coordinates follow as 
\begin{equation}
\frac{dx}{dS} = (1+\epsilon) \frac{dx}{ds} =  (1+\epsilon) \cos \theta, 
\quad \quad \frac{dy}{dS} = (1+\epsilon) \frac{dy}{ds}  = (1+\epsilon)\sin \theta.
\end{equation}
Introducing the normal $\vec n=-\sin\theta \vec{e}_x+\cos \theta \vec{e}_y$ and tangential vectors $\vec t(s)=\cos \theta(s) \vec{e}_x+\sin\theta(s) \vec{e}_y$, the kinematic connection can be summarised as 
\begin{equation}\label{eq:r}
\frac{d\vec r}{dS} = (1+\epsilon)\vec t, \quad \Rightarrow \quad 
\vec r(S) =\int_{-\infty}^{S} (1+\epsilon)\vec t \, dS',
\end{equation} 
where $\vec r(S)$ is the position vector of a point at the interface.
\begin{figure}[b!]
\includegraphics{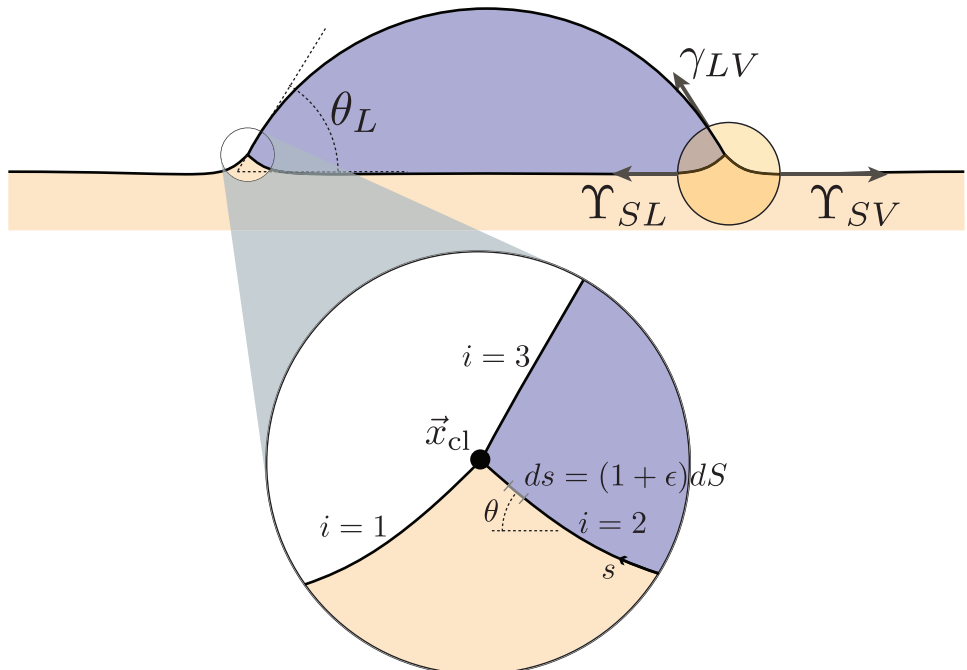}
\vspace{- 4 mm}
\caption{Definitions of the coordinate system used in the derivation.}
\label{schematics}
\vspace{- 5 mm}
\label{Fig1}
\end{figure}

\subsection{Elastic traction}
Consequently, the variations $\delta \epsilon(S)$ and $\delta \theta(S)$ displace points at the interface as 
\begin{equation}
\delta \vec r(S) = 
\int_{-\infty}^{S} \left[  \vec t \, \delta \epsilon +  (1+\epsilon)\vec n \, \delta \theta  \right]\, dS'.
\end{equation}
We will also define the associated displacement as a function of the deformed coordinate, as $\delta \vec u(s)\equiv \delta \vec r(S)$. With this we compute the variation of the elastic energy $\delta \mathcal F_e$, from the work done at the interface
\begin{equation}
\delta \mathcal F_e = \int_{-\infty}^r ds \, \vec \sigma(s) \cdot \delta \vec u(s) =   \int_{-\infty}^R dS \, \vec \Sigma(S) \cdot \delta \vec r(S).
\end{equation}
Here we introduced the true traction $\vec \sigma$ (force per unit area in the current state) and the nominal traction $\vec \Sigma$ (force per unit area in the reference state), which are related as $\vec \Sigma = (1+\epsilon)\vec \sigma$. Combining with the expression for $\delta \vec r(S)$, 

\begin{eqnarray}
\delta \mathcal F_e &=& 
\int_{-\infty}^R dS \, \vec \Sigma(S) \cdot \int_{-\infty}^{S} \left[  \vec t(S') \, \delta \epsilon(S') +  (1+\epsilon(S'))\vec n(S') \, \delta \theta(S')  \right]\, dS'.
\end{eqnarray}
Here it is important to keep track of the dependencies on $S$ and $S'$. Inverting the order of integration this becomes

\begin{eqnarray}\label{eq:pop}
\delta \mathcal F_e &=& 
\int_{-\infty}^R \, \left[ \vec t(S) \, \delta \epsilon(S) +   (1+\epsilon(S))\vec n(S) \, \delta \theta(S)  \right] dS 
\cdot \int_S^R \vec \Sigma(S') \,dS'.
\end{eqnarray}
From the definition of the functional derivatives

\begin{equation}
\delta \mathcal F_e = \int_{-\infty}^R \, \left[ \frac{\delta \mathcal F_e}{\delta \epsilon(S)} \, \delta \epsilon(S) +  
\frac{\delta \mathcal F_e}{\delta \epsilon(S)} \, \delta \theta(S)  \right] dS,
\end{equation}
the expression (\ref{eq:pop}), can be written as

\begin{equation}\label{eq:Gamma}
\vec \Gamma \equiv \frac{\delta \mathcal F_e}{\delta \epsilon(S)} \vec t + 
\frac{1}{1+\epsilon(S)} \frac{\delta \mathcal F_e}{\delta \theta(S)} \vec n =  \int_S^R \vec \Sigma(S') \,dS' 
=   \int_s^r \vec \sigma(s') \,ds'.
\end{equation}
Taking the derivative $d\vec \Gamma/ds$ gives Equation 3 of the paper.

\section{Free energy and constraints}

The sum of elastic and capillary free energies reads
 \begin{equation}
\mathcal{\tilde F} = \mathcal{F}_e + \sum_i \int_{-\infty}^{R_i} dS \, (1+\epsilon_i) \gamma_i(\epsilon_i).
 \end{equation}
Here we now explicitly derive the boundary conditions by introducing the constraints that were explained in the main text. The first is that the three interfaces meet at the same spatial point $\vec r_{\rm cl}$, as can be computed from the kinematic relation (\ref{eq:r}). The second constraint is on the material points $R_1+R_2$ is constant, so that $\delta R_1= -\delta R_2$. The total functional to be minimised, including the constraint using Lagrange multipliers $\lambda$ and $\vec f_i$, then becomes

\begin{eqnarray}\label{eq:functional}
\mathcal{F} &=&  \mathcal{F}_e + \sum_i \int_{-\infty}^{R_i} dS \, (1+\epsilon_i) \gamma_i(\epsilon_i) \, + \lambda(R_1+R_2) + \sum_{i} \vec f_i \cdot \left[\vec r_{cl} - \int_{-\infty}^{R_i} dS \, (1+\epsilon_i) \vec t(\theta_i) \right].
\end{eqnarray}
The Lagrange multipliers $\vec f_i$ will turn out to have a natural interpretation as contact line forces.

\subsection{Field equations and boundary conditions}

Variation of the functional (\ref{eq:functional}) with respect to $\epsilon_i(S)$ yields,
\begin{equation}\label{eq:depsilon}
 \frac{\delta {\cal F}}{\delta \epsilon_i(S)}= \vec \Gamma_i \cdot \vec t_i   + \Upsilon_i - \vec{f}_i \cdot \vec{t}_i = 0, \quad \textrm{all} \quad S < R_i,
\end{equation}
while the variation of $\theta_i(S)$ becomes,
\begin{equation}\label{eq:dtheta}
\frac{1}{1+\epsilon}\frac{\delta {\cal F}}{\delta \theta_i(S)} =\vec \Gamma_i \cdot \vec n_i  - \vec{f}_i \cdot \vec{n}_i = 0, \quad \textrm{all} \quad S < R_i.
\end{equation}
Here remind the definition of the integrated elastic traction $\vec \Gamma$ in (\ref{eq:Gamma}). Combining these two relations, the force can thus be written as a sum of elastic and capillary contributions

\begin{equation}\label{eq:eqS}
\vec{f_i} = \Upsilon_i \vec{t}_i + \vec \Gamma_i, \quad \textrm{all} \quad S < R_i.
\end{equation}
This result was already derived in the manuscript in differential form (equation 4). 

Our main interest here is the boundary condition. Variation of (\ref{eq:functional}) with respect to the contact line position $\delta \vec r_{\rm cl}$ gives 
\begin{equation}\label{eq:dxy}
\sum_i \! \vec{f}_i = \vec{0},
\end{equation}
which is interpreted as balance of forces. Evaluating (\ref{eq:eqS}) at the contact line, where $\vec \Gamma=0$, it follows that $\vec f_i= \Upsilon_i \vec t_i$ at the contact line. (\ref{eq:dxy}) then gives the Neumann condition, equation 5 in the paper. Finally, the no-pinning boundary condition is found from variation $\delta R_i$, which gives 
\begin{equation}\label{eq:dR}
\frac{\partial \mathcal{F}}{\partial R_i}=\frac{\partial \mathcal{F}_e}{\partial R_i} + (1+\epsilon_i) \gamma_i - (1+\epsilon_i) \vec{f}_i \cdot \vec{t}_i + \lambda (\delta_{i1} + \delta_{i2}) = 0, 
\quad \textrm{for} \quad S = R_i,
\end{equation}
Eliminating $\lambda$ from the equations for $i=1$ and $i=2$, we finally obtain (again using $\vec \Gamma =0$ at the boundary):
\begin{equation}
(1+\epsilon_1)^2 \gamma_1'  -\frac{\partial \mathcal{F}_e}{\partial R_1}  = (1+\epsilon_2)^2 \gamma_2'  - \frac{\partial \mathcal{F}_e}{\partial R_2}, \quad \quad \textrm{all} \quad S = R_i.
\end{equation}
This is the no-pinning condition equation 6. 

The only remaining step is to derive equation 9 in the manuscript. As usual, the derivation of Young's boundary condition is achieved from a global displacement of the contact line -- including the wetting ridge. This is achieved from the combined variation

\begin{eqnarray}
0= \frac{\partial {\mathcal F}}{\partial R_i}-\int_{-\infty}^{R_i} \left(\theta_i'(S)\frac{\delta {\mathcal F}}{\delta \theta_i(S)}+\epsilon_i'(S)\frac{\delta {\mathcal F}}{\delta \epsilon_i(S)}\right) dS = \nonumber \\
\left[ (1+\epsilon_i) ( \gamma_i+\vec \Gamma_i \cdot \vec t_i -  \vec f_i \cdot  \vec t_i ) \right]_{-\infty} + \lambda (\delta_{i1} + \delta_{i2}) +\frac{\partial \mathcal{F}_e}{\partial R_i}-\int_{-\infty}^{R_i}  (1+\epsilon_i)  \vec \Sigma_i \cdot \vec t_i dS,
\end{eqnarray}
where we made use of the equilibrium relations expressed above. Combined with equations 7 and 8 in the manuscript, we get the generalised Young's equation 9.

\section{Rheometry}
The aim of our home-build rheometer setup (inset of fig. 2) is to probe the response of the gel in the direction perpendicular to the direction of a uniaxial exernal strain $\epsilon_\infty$. A small magnetic disc (6 mm diameter) is laid on a a large gel sample large with typical thickness 5 mm. A coil carefully positioned above the magnet creates a vertical magnetic force $F$ proportional to the current in the coil. Force calibration is performed with the magnet sitting on a precision scale and operating the coil with DC current. The vertical position $Z$ of the magnet is measured with a laser vibrometer. The output of the vibrometer as well as the current in the coil are acquired simultaneously, yielding the complex effective stiffness $K \equiv F/Z$, transverse to the direction of the static stress.
\begin{figure}[t!]
\includegraphics{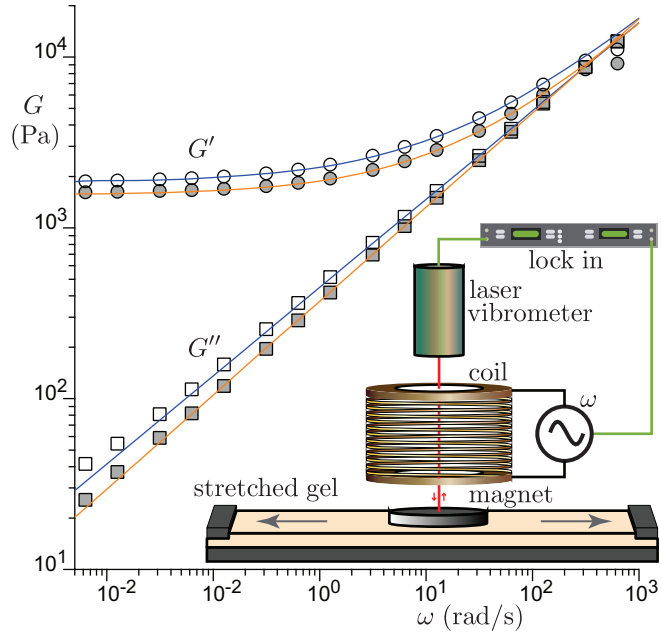}
\vspace{- 4 mm}
\caption{Storage $G'$ and loss $G''$ moduli of a gel sample as a function of  $\omega$. Closed and open symbols: un-strained and strained ($\epsilon_\infty=0.76$) samples respectively. Experimental data are well fitted by $G' + i G''= G\left[1+ (i \tau \omega)^n \right]$ ($n=0.55$ for $\epsilon_\infty=0$ and $n=0.51$ for $\epsilon_\infty=0.76$). \textit{Inset}: setup to determine the linear response of a soft solid transverse to the direction of uniaxial strain.}
\vspace{- 5 mm}
\label{rheo}
\end{figure}

\begin{figure}[h!]
		\centering
			\includegraphics{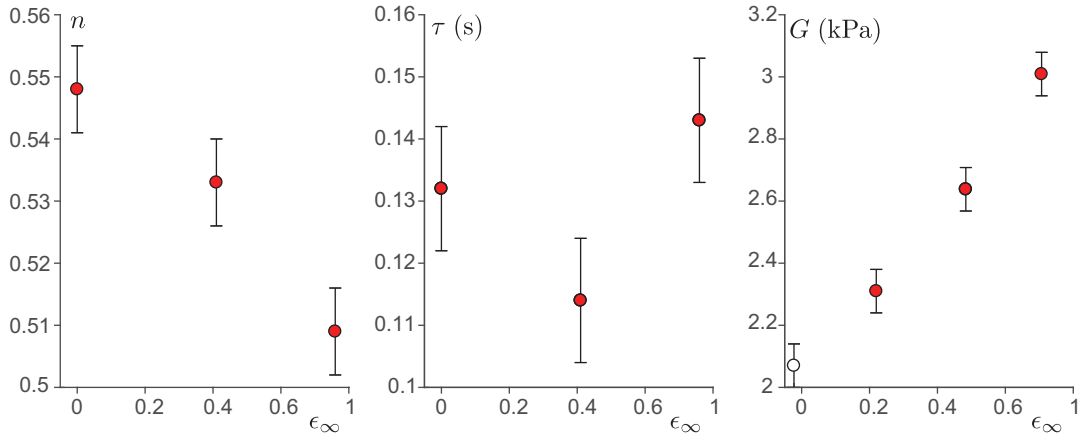}
		\caption{Gel rheology as a function of the strain $\epsilon_\infty$. The shear modulus $G$ is of the form: $G(\omega)=G(0)\left[1+(i\omega\tau)^n\right]$. From left to right: parameters $n$, $\tau$ and $G(0)$ as a function of  $\epsilon_\infty$.}
		\label{RheoSup}
	\end{figure}

We use this rheometer to determine the frequency dependence of $K(\omega)$, which is identical to that of the Young's or shear modulus. We find that  $K$ can be fitted by the simple form : $K(\omega)=K(0)\left[1+(i\omega\tau)^n\right]$ for  $\epsilon_\infty$ up to one $K$. The frequency dependence varies very little with $\epsilon_\infty$, as shown in Fig \ref{RheoSup}. The exponent $n$ decreases slightly from 0.55 down to 0.50 for $\epsilon_\infty$ increasing from 0 to 1, and the time constant $\tau$ is constant within experimental uncertainty ($\tau = 0.13$ s). The values of $n$ and $\tau$ for zero strain are in perfect agreement with previous rheological studies on the same gel.\\

The analytical relation between the stiffness at zero frequency $K_0$ and the shear modulus $G(0)$ is known  only for an infinite thickness. In this case $K_\infty=G(0)d/(1-\nu)$, where $d$ is the disc diameter and $\nu$ the Poisson's ratio (we took $\nu=0.5$ \cite{nu}).  Thus, $G(0)$ was obtained through  complementary static indentation measurements with indentors much smaller that $e$. As shown in Fig. \ref{RheoSup}, we find again that $G(0)$ is weakly dependent on $\epsilon_\infty$. In the end, the data $G(\omega)$ shown in fig. 2 are obtained through: $G(\omega)=G(0)K(\omega)/K(0)$.

%
\bibliographystyle{apsrev4-1}

%

\end{document}